# Solar wind parameters in rising phase of solar cycle 25

Yuri I. Yermolaev, Irina G. Lodkina, Alexander A. Khokhlachev, Michael Yu. Yermolaev, Maria O. Riazantseva, Liudmila S. Rakhmanova, Natalia L. Borodkova, Olga V. Sapunova, Anastasiia V. Moskaleva

Space Research Institute, Russian Academy of Sciences, 117997 Moscow, Russia

**Abstract**

Solar activity and solar wind parameters decreased significantly in solar cycles (SCs) 23–24. In this paper, we analyze solar wind measurements at the rising phase of SC 25 and compare them with similar data from the previous cycles. For this purpose, we simultaneously selected the OMNI database data for 1976-2022, both by phases of the 11-year solar cycle and by large-scale solar wind types (in accordance with catalog http://www.iki.rssi.ru/pub/omni), and calculated the mean values of the parameters for the selected datasets. The obtained results testify in favor of the hypothesis that the continuation of this cycle will be similar to the previous cycle 24, i.e. SC 25 will be weaker than SCs 21 and 22.

**1. Introduction**

The solar wind (SW), which is formed by the expansion of the hot solar corona into the interplanetary medium, is one of the main goals of space research. On the one hand, the study of the solar wind makes it possible to better understand the properties of the solar atmosphere and the processes of plasma outflow from it [1,2,3]. On the other hand, the solar wind is the main agent that carries disturbances from the Sun to the Earth and causes space weather effects [4, 5, 6].

Direct measurements of the solar wind began at the beginning of the space age [7, 8] and cover solar cycles 20-25 (see, for example, base of solar wind measurements https://spdf.gsfc.nasa.gov/pub/data/omni /low_res_omni [9]). The beginning of this period fell on the epoch of high solar activity, and at the minimum between solar cycles 22 and 23, a decrease in solar activity began, which continued during SC 23 and 24 [10, 11, 12]. The decrease in solar activity was accompanied by a number of significant changes in the solar wind and its impact on the Earth's magnetosphere [13, 14, 15,16,17,18, 19]: (1) a change in the structure heliosphere (for example, a decrease in the number of CMEs and their manifestations in the interplanetary medium with an almost unchanged number of high-speed streams from coronal holes and associated CIR compression regions), (2) a decrease in solar wind parameters, both in different phases of the solar cycle and in different types of solar wind streams, and (3) a decrease in the disturbance of the magnetosphere, in particular, a decrease in the number of magnetic storms on Earth by almost 10 times. At present, the Sun has passed the rising phase of the 25th solar cycle (see the behavior of the average annual values of sunspots in the period 2019-2022 in Fig. 1), and direct measurements of the solar wind for this phase are available for research. Together with solar observations, the analysis of these measurements makes it possible to verify models that predict the development of the current solar cycle and, in particular, to obtain more reliable predictions of the behavior of the Sun, the heliosphere, and space weather effects near the solar cycle maximum [20, 21, 22, 23].

In our previous article [15], we analyzed how the average parameters changed in various large-scale solar wind streams at different phases of 21-24 solar cycles (1976-2019). For this purpose, we simultaneously selected the data from the OMNI database [9], both by solar cycle phases and by large-scale solar wind types [24], and calculated the average values of the parameters for the selected data sets. As a result, it was shown that in SC 23-24 (1997-2019) for the corresponding phases of solar cycles for all types of solar wind streams, the parameters decreased by 20-40% compared to SC 21-22. In this work, a similar selection of the OMNI database data for the rising phase of 25 SCs is carried out and, for the first time, a comparison is made with similar phases of 4 previous SCs in order to determine the similarity and difference between the current SC and previous SCs and predict its development.

## 2. Data and Methods

In this work, we use the same sources of information as in the previous paper [15]: (1) hourly data of the OMNI database on solar wind measurements for 1976–2022 (https://spdf.gsfc.nasa.gov/pub/data/omni/low_res_omni [9], accessed on 1 March 2023), and (2) intervals of different SW types in a living catalog of large-scale phenomena since 1976 (http://www.iki.rssi.ru/pub/omni [24], accessed on 1 March 2023), created on the basis of the OMNI database.

In accordance with the catalog, the following large-scale (>$10^6$ km) solar wind types were identified: quasi-stationary types (1) Heliospheric current sheet, HCS, (2) slow streams from the region of coronal streamers, Slow, (3) fast streams from the region of coronal holes, Fast, and perturbed types (4) compression regions between slow and fast flow types - corotating interaction regions, CIR, (5) compression regions between slow flow type and fast manifestations ICME, Sheath, and (6, 7) 2 variants of ICME - Ejecta and magnetic cloud (MC), which are distinguished by a higher and more regular interplanetary magnetic field (IMF) in MC compared to Ejecta. The classification of solar wind phenomena used is generally accepted (see [15] for more details), the type identification method is described in detail in paper [24].

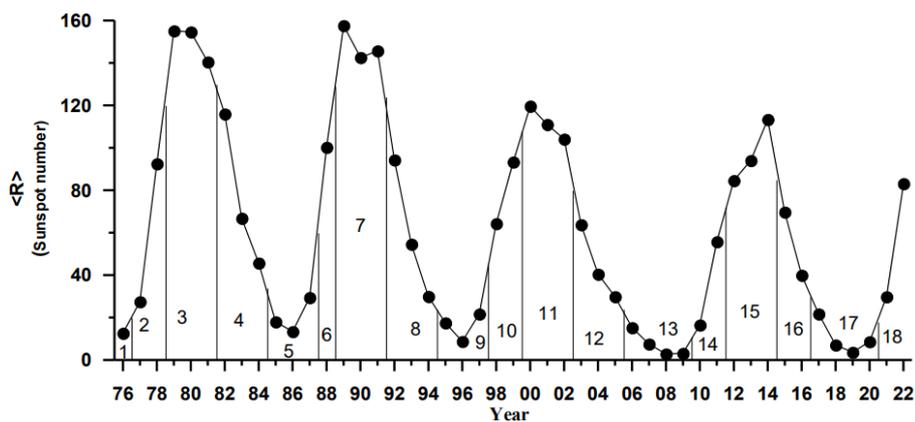

**Figure 1.** *Annual data of the sunspot number (The numbers and vertical lines show the division over phases of solar cycles 21–25.).*

Table 1. Averaging intervals over phases of solar cycle

| № interval | № Cycle | Phase of solar cycle | Years |
|---|---|---|---|
| 1 | 21 | minimum phase | 1976 |
| 2 | | rising phase | 1977,1978 |
| 3 | | maximum phase | 1979-1981 |
| 4 | | declining phase | 1982-1984 |
| 5 | | minimum phase | 1985-1987 |
| 6 | 22 | rising phase | 1988 |
| 7 | | maximum phase | 1989-1991 |
| 8 | | declining phase | 1992-1994 |
| 9 | | minimum phase | 1995-1997 |
| 10 | 23 | rising phase | 1998-1999 |
| 11 | | maximum phase | 2000-2002 |
| 12 | | declining phase | 2003-2005 |
| 13 | | minimum phase | 2006-2009 |
| 14 | 24 | rising phase | 2010,2011 |
| 15 | | maximum phase | 2012-2014 |
| 16 | | declining phase | 2015-2016 |
| 17 | | minimum phase | 2017-2020 |
| 18 | 25 | rising phase | 2021-2022 |

The entire time interval 1976-2022 was divided into 18 sub-intervals corresponding to the phases of 21-25 solar cycles (see Fig. 1 and Table 1). In contrast to previous works [15,16], in this work the phase of the 24/25 SC minimum is extended and includes the period 2017-2020, and in addition, the period of the rising phase 2021-2022 is added. In each of the 18 sub-intervals and for each of the 8 types of SW (the 7 listed above + their sum), the data were averaged. All parameters in the averaging intervals have a large statistical spread and their standard deviation is close to the average value. However, due to the large (~$10^3$) number of points in the averaging sets for all types of SW (except MC, where the statistics are small [15]), the statistical error (i.e., the standard deviation divided by the square root of the number of measurement points) turns out to be small, and these trends in the behavior of the parameters have sufficient statistical significance [25]. It should be noted that the largest scatter of parameters is observed for the proton temperature T, and since it has a lognormal distribution [26, 27], we averaged the logT value.

## 3. Results

Figures 2-7 present time profiles of parameters of solar wind plasma and interplanetary magnetic field (IMF) averaged over phases of solar cycles (Table 1): minimum – black circles, rising phase – blue triangles, maximum – purple squares, declining phase – green inverted triangles, without selection with phases – red open squares. Right dots (blue triangles) in all panels correspond to the rising phase of 25 SC and these values represent the main result of this work. The data for magnetic clouds are widely scattered in all figures due to the small number of events.

We will start the analysis with the solar wind bulk velocity, which, as shown earlier [15], turned out to be the least affected by the weakening of solar activity in 23-24 SCs. Figure 2 shows that the solar wind velocity is quite stable, and it weakly depends on both the number and phase of the SC, and the type of SW. The only short-term increase in the average velocity was observed for events generated by CMEs (Sheath, Ejecta and MC) during the declining phase of SC 23 and is associated with a short-term increase in solar activity (in particular, the extreme events of October-November 2003 and 2004). The deviation of the average velocity at the rising phase of SC 25 from the previous minimum phase for different types of SW is less than 20 km/s, this value is noticeably lower than the standard deviations and corresponds to the velocity behavior at the beginning of SC 24. Thus, no specific features are observed in the rising phase of SC 25 compared to the rising phase of SC 24.

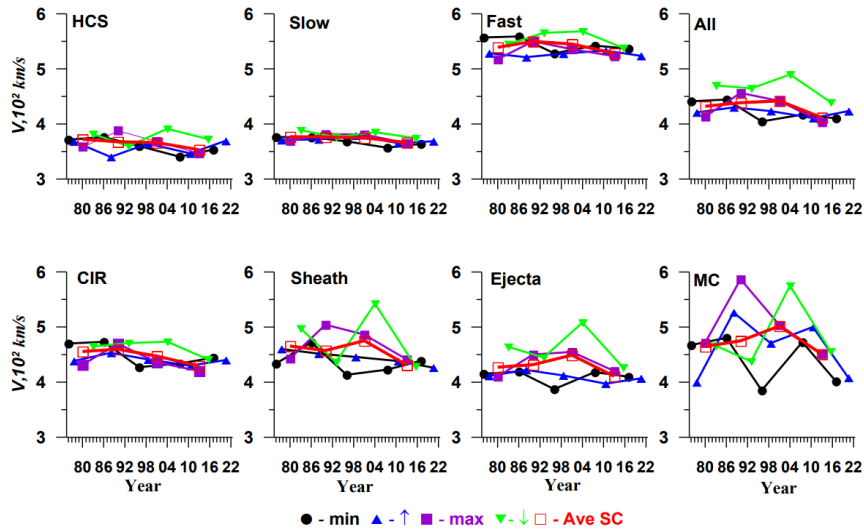

**Figure 2.** *Time profiles of bulk velocity V in 7 different types of SW (heliospheric current sheet – HCS, Slow and Fast streams, CIR, Sheath, Ejecta and MC) and without SW type selection (All)*

Figure 3 shows the variations in the logarithm of the proton temperature, logT. Despite the large temperature scatter, profiles averaged over the phases of solar cycles on a logarithmic scale have fairly smooth shapes with a pronounced tendency for temperature increase in the phases of maximum and decline of SC 22-24. It can be noted that in quasi-stationary types of SW (HCS, Slow and Fast) and without selection by SW types (panel All), the temperature at the rising phase of 25 SC has a weak tendency to increase compared to both the phase of the previous minimum and the rising phase of 24 SC. Such a trend is not observed for disturbed types of SW.

Figure 4 represents density variations, N. Despite the large scatter of values (slightly smaller than the temperature scatter), the curves for quasi-stationary types of SW are quite smooth and show a tendency to higher density values during the minimum and declining phases. The density after the minimum of 22/23 SC drops noticeably in all SC phases and for all types of SW. Comparison of the dynamics of density at the transition from the phase of minimum to the rising phase of 25 SC shows a behavior similar to the previous 24 SC.

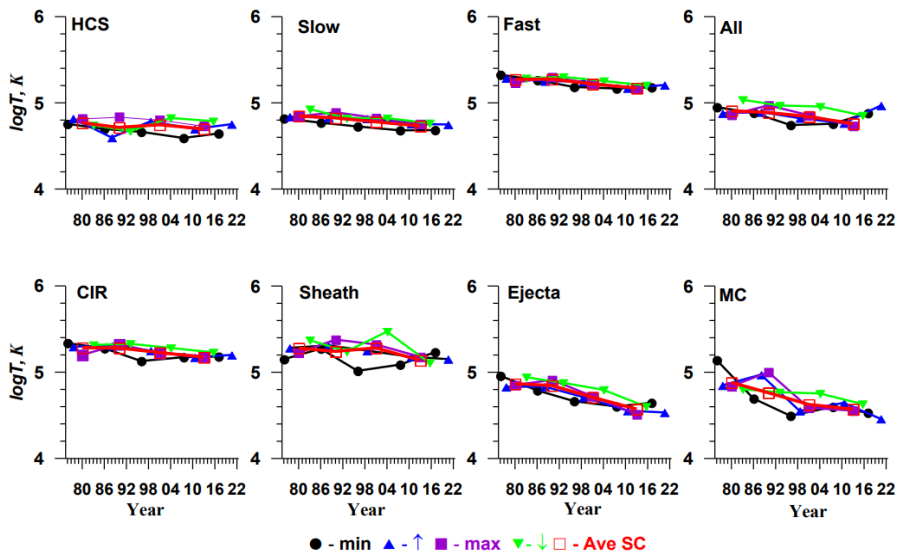

**Figure 3.** *Time profiles of logarithm of proton temperature T*

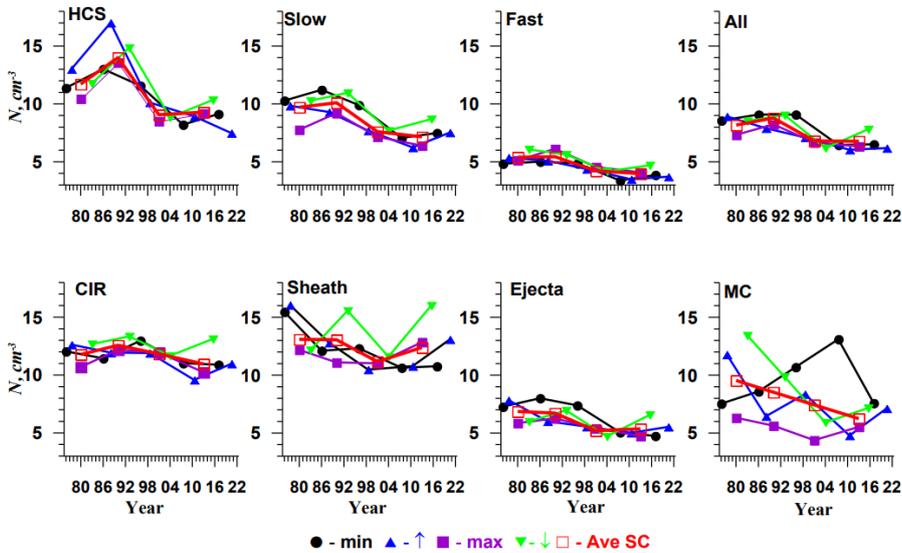

**Figure 4.** *Time profiles of density N*

Time profiles of the magnitude of interplanetary magnetic field B are presented in Figure 5. For different types of SW, the curves show higher values of B for the phases of maximum and decline, and after the minimum of 22/23 SC, a decrease in the magnetic field is observed. For quasi-stationary SW (HCS, Slow and Fast) and without selection by SW types (panel All), the parameter B remains unchanged or slightly increases compared to the previous minimum phase and rising phase of 24 SC.

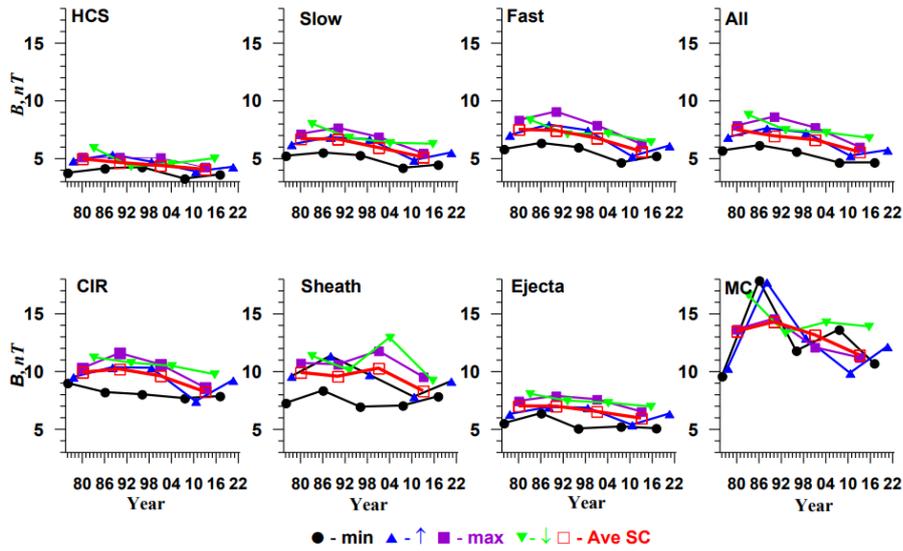

**Figure 5.** *Time profiles of magnitude of IMF B*

It is interesting to compare the time profiles of the dimensionless value beta, the ratio of the thermal pressure of protons to the magnetic pressure (Figure 6). For all types of SW, in contrast to the previous figures, the beta parameter increases at the minimum phases and decreases at the maximum phases and demonstrates a slight decrease in the epoch of solar activity decrease at 23-24 SCs. In the rising phase of the 25 SC, the beta parameter behaves similarly to the previous 24 SC.

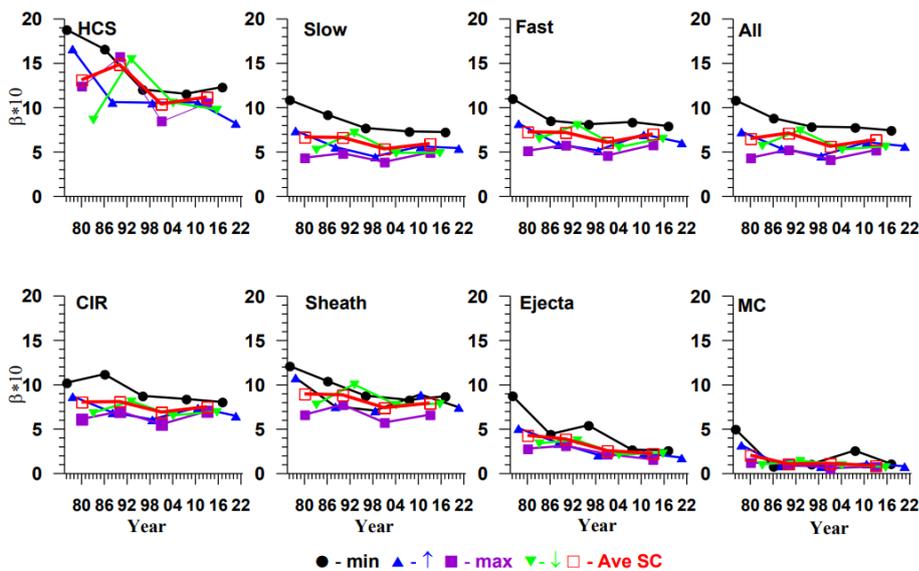

**Figure 6.** *Time profiles of proton β-parameter*

The helium abundance (relative density of alpha particles) Nα/Np is presented in figure 7, which shows that for all types of SW parameter Nα/Np is maximum at the phases of maximum and minimum at the phases of minimum, and in the epoch of low solar activity at 23-24 SCs it dropped ~1.5 times. It is important to note that if the proton density dropped by ~40%, then the absolute density of alpha particles dropped by a factor of ~2. For all types of SW, the Nα/Np increases in comparison with the phase of the previous minimum and behaves similarly to the rising phase of 24 SC.

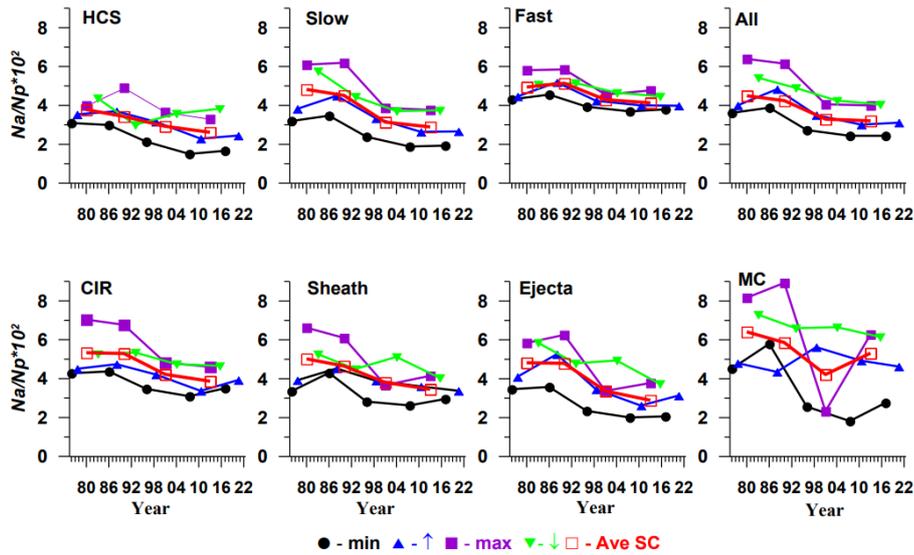

**Figure 7.** *Time profiles of helium abundance Nα/Np*

## 4. Discussion and Conclusions

In this paper, we simultaneously selected the OMNI database data [9] for 1976-2022, both by phases of the 11-year solar cycle and by large-scale solar wind types from catalog http://www.iki.rssi.ru/pub/omni [24], and calculated mean parameter values for the selected datasets. In contrast to the previous work [15], for the first time in this way we calculated the average values of the parameters for the rising phase of SC 25 and compared them with similar phases of previous cycles of low solar activity 23-24. This analysis shows that there are no significant reasons to believe that the beginning of the current 25 SC differs from the previous cycle, and most likely the continuation of this cycle will be similar to the previous cycle 24, i.e. 25 SC will be weaker than 21 and 22 SCs.

Our definition of the time limits of the rising phase of 25 SC (this is difficult to do for a cycle that has not yet ended) is rather arbitrary. Given the above conclusion about a weak 25 SC, it cannot be ruled out that the 2021-2022 interval includes measurements related to the maximum phase of the 25 SC. However, their inclusion in our analysis would only strengthen these trends, while the dependences obtained were so weak that they did not allow us to conclude that the rising phase of 25 SC differs from the analogous phase of the previous 24 SC. Therefore, the boundaries of the rising phase of 25 SC adopted in this study, which possibly include partially the maximum phase, cannot affect the conclusions drawn.

The prediction of the development of solar activity in the coming years and, in particular, in 25 SC remains debatable and is widely discussed in the specialized literature [23, 28, 29, 30, 31, 32, 33]. We hope that the results presented in this paper will contribute to this discussion and shed additional light on the development of solar activity in the current solar cycle and beyond.

**Author Contributions**

Conceptualization, Y.I.Y.; methodology, Y.I.Y.; software, I.G.L. and A.A.K.; validation, M.Y.Y., M.O.R. and L.S.R.; formal analysis, N.L.B.; investigation, O.V.S.; resources, A.V.M.; data curation, I.G.L.; writing—original draft preparation, Y.I.Y.; writing—review and editing, Y.I.Y.; visualization, I.G.L.; supervision, Y.I.Y. All authors have read and agreed to the published version of the manuscript.

**Funding**

The work was supported by the Russian Science Foundation, grant 22-12-00227.

**Acknowledgments**

Authors thank creators of databases **https://spdf.gsfc.nasa.gov/pub/data/omni/low_res_omni** (1 March 2023) and **http://www.iki.rssi.ru/pub/omni** (1 March 2023) for possibility to use in the work.

**Conflicts of Interest**

The authors declare no conflict of interest.